\input epsf.tex

\documentclass[preprint,showpacs,preprintnumbers,amsmath,amssymb]{revtex4}

\usepackage{graphicx}
\usepackage{dcolumn}
\usepackage{bm}

\begin{document}


\title{Partial Decoherence of Histories and the Di\'osi Test}

\author{J.J.Halliwell}%
\affiliation{Blackett Laboratory \\ Imperial College \\ London SW7
2BZ \\ UK }



\begin{abstract}
In the decoherent histories approach to quantum theory, attention
focuses on the conditions under which probabilities may be
assigned to sets of quantum histories. A variety of conditions have been
proposed, but the most important one is decoherence, which means
that the interference between every pair of histories in the set is zero.
Weaker conditions have been considered, such as consistency, or
linear positivity, but these are ruled out by the requirement of
consistent composition of subsystems, proposed by Di\'osi. Here we
propose a new condition which we call {\it partial decoherence},
and is the requirement that every history has zero interference
with its negation. This is weaker than decoherence and stronger
than linear positivity (but its relation to consistency is less
simply defined -- it is neither stronger nor weaker).
Most importantly, it satisfies the Di\'osi condition.
A strengthened Di\'osi condition is proposed, which partial decoherence
narrowly fails, due to an unusual property of inhomogeneous histories.
In an appendix an example is given of a set of histories which
are consistent but not decoherent.

\end{abstract}



\maketitle

\newcommand\beq{\begin{equation}}
\newcommand\eeq{\end{equation}}
\newcommand\bea{\begin{eqnarray}}
\newcommand\eea{\end{eqnarray}}

\def\A{{\cal A}}
\def\D{\Delta}
\def\H{{\cal H}}
\def\E{{\cal E}}
\def\p{\partial}
\def\la{\langle}
\def\ra{\rangle}
\def\ria{\rightarrow}
\def\x{{\bf x}}
\def\y{{\bf y}}
\def\k{{\bf k}}
\def\q{{\bf q}}
\def\p{{\bf p}}
\def\P{{\bf P}}
\def\r{{\bf r}}
\def\s{{\sigma}}
\def\a{\alpha}
\def\b{\beta}
\def\e{\epsilon}
\def\U{\Upsilon}
\def\G{\Gamma}
\def\om{{\omega}}
\def\Tr{{\rm Tr}}
\def\ih{{ \frac {i} { \hbar} }}
\def\trho{{\rho}}

\def\au{{\underline \alpha}}
\def\bu{{\underline \beta}}
\def\pp{{\prime\prime}}

\def\id{{1 \!\! 1 }}
\def\half{\frac {1} {2}}

\def\jjh{j.halliwell@ic.ac.uk}

\section{Introduction}

The decoherent histories approach to quantum theory has proved to be
a very useful viewpoint from which to address the emergence of classical behaviour
from quantum theory and also for analyzing the conceptual structure of quantum
theory itself \cite{GH1,GH2,Gri,Omn1,Omn2,Hal2,Hal3,Hal4,DoK,Ish}. The central idea is to determine the conditions under which
probabilities may be assigned to histories of a closed system and then to
examine the predictions of those probabilities. A variety of different
probability assignment conditions have been proposed, of differing strengths
and mathematical consequences.

A significant step in discriminating between these
conditions
was made by Di\'osi, who proposed that any such condition should satisfy certain
reasonable requirements of statistical independence when applied
to composite systems consisting of non-interacting independent subsystems \cite{Dio}.
This reduced the number of different
probability assignment conditions to just one,
namely diagonality of the decoherence functional, a condition we
will refer to as decoherence of histories  (or more simply, decoherence).
For a pure initial
state, this is equivalent to demanding that
the states corresponding to each history should be orthogonal.

The work described in this paper arose as a result of the realization
that it is possible to weaken the condition of decoherence and still
pass the Di\'osi test. This weakened condition is called partial decoherence
and for a pure initial state it is the requirement that the state for each
history is orthogonal to the state representing the negation of that history.
However, this new condition invites a revisiting of the Di\'osi test and
a strengthened version of the test is considered. Partial decoherence
narrowly fails to pass this strengthened test so is ultimately unsatisfactory.
These considerations underscore decoherence of histories as the most important
(and possibly only) viable condition for the assignment of probabilities
to histories and we discuss the physical reasons why this is the case.

At the encouragement of the editors, this paper is a speculative exploration
of ideas in progress, rather than a report of significant new results, so may come across
as incomplete in some parts.

\section{The Decoherent Histories Approach}

In quantum theory, alternatives at each moment of time are
represented by a set of projection operators $\{ P_a \}$,
satisfying the conditions
\bea
\sum_a P_a &=& 1
\\
P_a P_b &=& \delta_{ab} P_a
\eea
where we take $a$ to run over some finite range.
In the decoherent histories approach to quantum theory \cite{GH1,GH2,Gri,Omn1,Omn2,Hal2,Hal3,Hal4,DoK,Ish},
the simplest type
of history, a homogenous history, is represented by a class operator $C_{\a}$
which is a time-ordered string
of projections
\beq
C_{\a} = P_{a_n} (t_n) \cdots P_{a_1} (t_1)
\label{1.3}
\eeq
Here the projections are in the Heisenberg picture and $ \a $ denotes
the string $ (a_1, \cdots a_n)$. We take $\a$ to run over $N$ values so there
are $N$ histories.
The class operator Eq.(\ref{1.3}) satisfies the conditions
\beq
\sum_{\a} C_{\a} = 1
\label{1.4}
\eeq
and also
\beq
\sum_{\a} C^{\dag}_{\a} C_{\a} = 1
\label{1.5}
\eeq
Probabilities are assigned to histories via the formula
\beq
p(\a) = {\rm Tr} \left( C_{\a} \rho C_{\a}^{\dag} \right)
\label{1.6}
\eeq
which is in essence the usual Born rule generalized to histories.
These probabilities are clearly positive and normalized
\beq
\sum_{\a} p(\a ) = 1
\label{1.6b}
\eeq
which follows from Eq.(\ref{1.5}).

It is also natural to consider more complicated histories which are
given by {\it sums} of strings of projection operators of the
form Eq.(\ref{1.3}). These are called inhomogenous histories and typically
do not satisfy Eq.(\ref{1.5}), so their probabilities do not sum to $1$
in general
\beq
\sum_{\a} p(\a) \ne 1
\label{1.7}
\eeq
and in fact the individual probabilities may be greater than $1$ \cite{IshLin}.
(But the probabilities do sum to $1$ when there is decoherence, discussed below).
This difference between homogeneous and inhomogeneous
histories turns out to be important in what follows.

As the double slit experiment indicates, the assignment
of probabilities to histories in quantum mechanics is not always
possible. For the $p (\a) $ to be true probabilities
the histories must satisfy certain conditions which, loosely speaking,
ensure that there are no interference effects.
To this end, we introduce the decoherence functional
\beq
D(\a, \a') = {\rm Tr} \left( C_{\a} \rho C_{\a'}^{\dag} \right)
\eeq
which may be thought of as a measure of interference between pairs of histories.
It satisfies the conditions
\bea
D(\a, \a') &=& D^* (\a', \a)
\\
\sum_{\a} \sum_{\a'} D(\a, \a') &=& 1
\label{1.10}
\eea
and note that the probabilities are given by its diagonal elements
\beq
p(\a) = D(\a, \a)
\eeq

The simplest and most important condition normally
imposed is that the probabilities should satisfy the probability sum rules,
that is,
that they are additive for all disjoint pairs of histories.
More precisely, the probability of history $ \a$ or history $\a'$
must be the sum of $p(\a)$ and $p(\a')$. Since this combination
of histories is represented by the class operator $C_{\a} + C_{\a'}$, Eq.(\ref{1.6})
implies that
\beq
p( \a \ {\rm or} \  \a') = p(\a) + p(\a') + 2 \ {\rm Re} D(\a, \a')
\eeq
Hence for the probabilities to satisfy the expected sum rules
we require that
\beq
{\rm Re} D(\a, \a') = 0, \ \ \ \a  \ne \a'
\label{1.13}
\eeq
for all pairs of histories $\a, \a'$. This condition is called
{\it consistency} of histories, and if there are $N$ histories there
are $ \half N ( N-1 )$ such conditions. (The numbers of this and similar conditions
below are given for inhomogenous histories. Homogenous histories satisfy
Eqs.(\ref{1.5}), (\ref{1.6b}) which means that some of the conditions
will be satisfied identically).
Consistency of histories
ensures that the
probabilities defined by Eq.(\ref{1.6}) satisfy all the conditions
one would expect of a probability for histories.

In many practical situations, there is present a physical mechanism
(such as coupling to an environment) which causes Eq.(\ref{1.13})
to be satisfied, at least approximately, and in such situations, it is
typically observed that the imaginary part of the off-diagonal
terms of $ D(\a, \a')$ vanish as well as the real part.
It is therefore of interest to consider
the stronger condition of {\it decoherence}, which is
\beq
D(\a, \a') = 0, \ \ \ \a \ne \a'
\eeq
Since $D(\a, \a')$ is complex there are $ N ( N-1)$ such conditions.

This stronger condition is related to the existence of records \cite{GH2,Hal4}.
For a pure initial state it means that we can add an extra projection
operator $R_{\gamma}$ at the end of the history which is perfectly
correlated with the alternatives at earlier times. That is
\beq
{\rm Tr} \left( R_{\gamma} C_{\a} \rho C_{\a'}^{\dag} \right)
= \delta_{\gamma \a} \delta_{\gamma \a'} p (\a)
\eeq
which implies that
\beq
p(\a) = {\rm Tr} \left( R_{\a} \rho \right)
\eeq
so that the probabilities for histories reduce entirely to a projection
at a single moment of time. This corresponds to the idea that there
exists a record at fixed moment of time somewhere in the system,
like a photographic plate, which carries complete information about the
entire history of the system.

There appear to be very few examples of situations where the histories
are consistent but not decoherent (but one simple example is given in the
Appendix). This another reason why it is natural
to impose the requirement of decoherence.

It is now useful to define the quasi-probability,
\beq
q(\a) = {\rm Tr} \left( C_{\a} \rho \right)
\label{1.17}
\eeq
Because it is linear in the $C_{\a}$, this quantity sums to $1$ and
also satisfies the
probability sum rules, but it is not in general a real number.
However, it is closely related to the probabilities Eq.(\ref{1.6}), because
Eq.(\ref{1.4}) implies that
\bea
q (\a ) &=& {\rm Tr} \left( C_{\a} \rho C_{\a}^{\dag} \right) + {\rm Tr} \left( C_{\a} \rho
{\bar C}_{\a}^{\dag} \right)
\nonumber \\
&=& p(\a) + D(\a, {\bar \a})
\label{1.18}
\eea
Here $ {\bar C}_{\a} $ denotes the negation of the history $C_{\a}$,
\beq
{\bar C}_{\a} =1 - C_{\a} = \sum_{\b, \b \ne \a } C_{\b}
\label{1.19}
\eeq
This means that when there is decoherence we have that the probabilities
are given by the simpler expression
\beq
p(\a ) = q (\a)
\label{1.20}
\eeq
Decoherence therefore ensures that $q(\a)$ is real and positive, even though it
is not in general.

These properties of $q(\a)$ inspired Goldstein and Page \cite{GoPa} to suggest a formulation
of quantum theory in which the
probabilities are given by ${\rm Re} \ q(\a)$, subject only to the requirement
that
\beq
{\rm Re} \ q( \a) \ge 0
\label{1.22}
\eeq
a condition they refer to as {\it linear positivity}.
These clearly agree with the usual assignments $p(\a)$ when there is consistency,
but this condition is weaker than consistency so the reverse is not true.

These three conditions -- decoherence, consistency and linear positivity -- are
the main probability assignment conditions that have been discussed in the
literature. However, this is not an exhaustive list. Hartle, for example, has
discussed probability assignments in terms of the possibility of
settling bets \cite{Har}. Also, a number of different versions of the
condition of decoherence exist \cite{GH3}. We will not pursue these developments
here.

\section{Partial Decoherence -- A New Condition}

The first aim of this paper is to note that there is in fact a fourth condition for the assignment of
probabilities which is weaker than decoherence, but stronger than linear positivity.
It is neither stronger nor weaker than consistency.

Consider again steps Eq.(\ref{1.17})-(\ref{1.20}) which relate the $p(\a)$ and the $q (\a)$.
The proposed new condition is to require that
the histories satisfy Eq.(\ref{1.20}) for all $\a$. Since $q(\a)$
is complex in general the condition may be written
\bea
{\rm Re} \ q(\a) &=& p(\a)
\\
{\rm Im} \ q(\a) &=& 0
\eea
so there are $2 N$ conditions.
From Eq.(\ref{1.18}), it is equivalent to the condition
\beq
{\rm Tr} \left( C_{\a} \rho ( 1 - C_{\a}^\dag ) \right) = 0
\label{1.28}
\eeq
for all $\a$. This means that every history has zero interference with its negation, but there will
still be pairs of histories whose decoherence functional is non-zero in its off-diagonal terms.
It is therefore natural to call this new condition {\it partial decoherence}.
Partial decoherence is clearly weaker than decoherence. It is stronger than linear
positivity, since $q(\a)$ is explicitly set
to equal a real, positive number.

The relationship of partial decoherence to consistency is more complicated.
Partial decoherence, Eq.(\ref{1.28}), allows some of the off-diagonal parts of $ {\rm Re} D (\a, \a') $
to be non-zero, so is weaker than consistency in this respect.
On the other hand, Eq.(\ref{1.28}) requires the
imaginary parts of $ D(\a, \bar \a) $ to vanish, which is not implied by consistency,
so in this respect is stronger than consistency. So partial decoherence and consistency are different
conditions and neither implies the other. A given set of histories may satisfy one condition,
or the other, or both, or neither.

The logical relationships between the four conditions -- decoherence, partial decoherence, consistency
and linear positivity -- is represented in Figure 1.

\begin{center}
\epsfxsize=100mm
\epsfbox{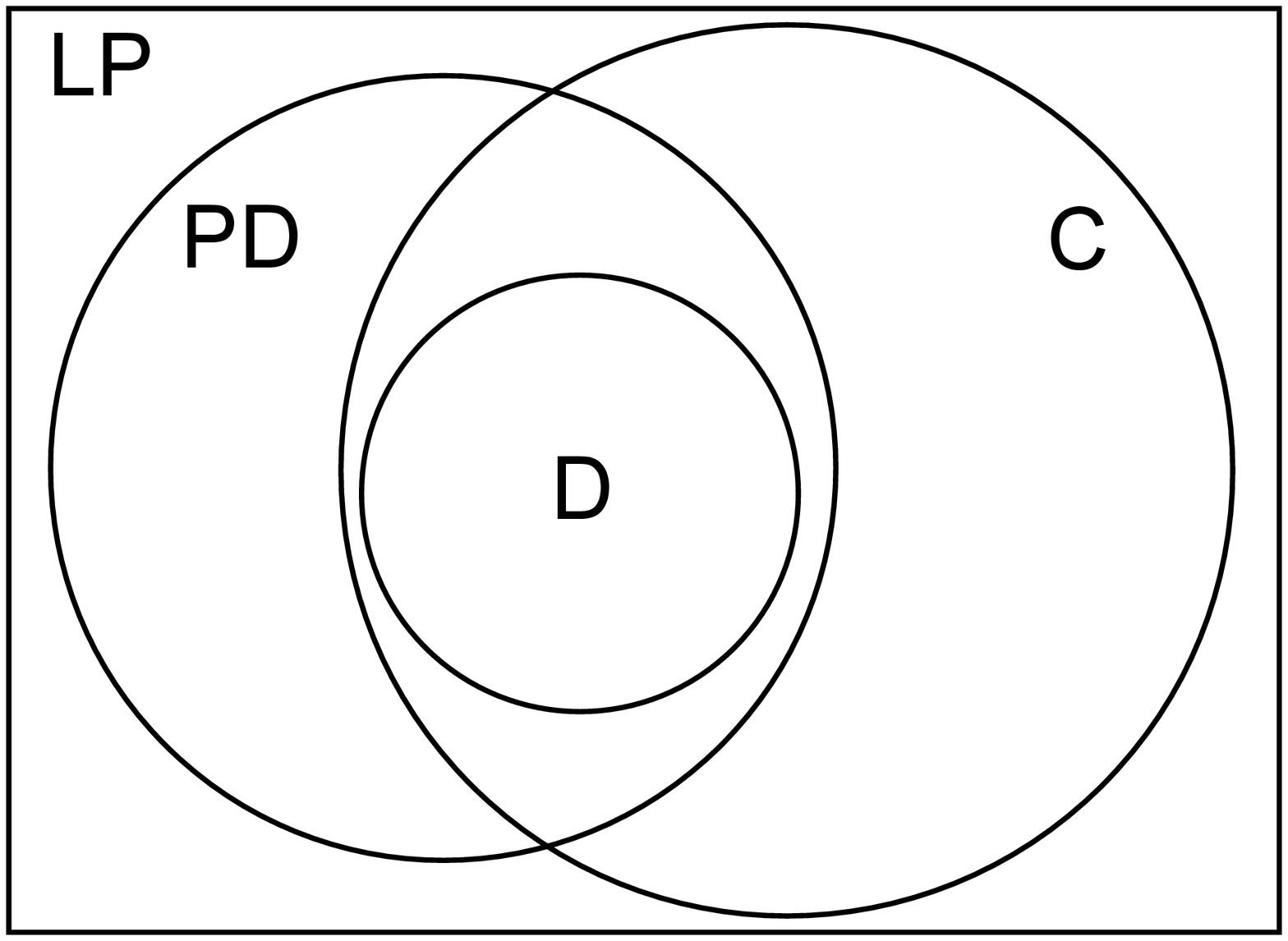}

\end{center}

\noindent{\bf Figure 1}: {\it A Venn diagram showing the relationships between
sets of histories satisfying
the four conditions of decoherence (D), partial decoherence (PD), consistency (C)
and linear positivity (LP).}

Note also that all four conditions may be
written as simple restrictions on the decoherence functional, and for conciseness, all four are
listed below in the order decoherence, partial decoherence, consistency
and linear positivity:
\bea
D(\a, \a' ) &=& 0, \ \ {\rm if} \ \ \a \ne \a'
\\
\sum_{\a', \a' \ne \a} D( \a, \a' ) &=& 0
\\
{\rm Re} D (\a, \a') &=& 0 \ \ {\rm if} \ \ \a \ne \a'
\\
\sum_{\a'} {\rm Re} D(\a, \a') & \ge & 0
\eea

\section{The Di\'osi Test}

This multitude of conditions raises the question, which of them are necessary for
a consistent formulation of the quantum theory of histories?
An important step in answering this question was made
by Di\'osi \cite{Dio}. He proposed the reasonable requirement that any assignment of probabilities
must behave in a reasonable way for statistically independent subsystems.
More precisely,

\noindent{\bf The Di\'osi Test:} {\it Suppose a composite system consists of a number of statistically
independent subsystems. Then any condition for the assignment of the probabilities
applied to the subsystems must imply the same condition for
the composite system.}

Suppose we have two independent subsystems $A$ and $B$, with state $ \rho^A \otimes \rho^B$,
and histories with class operators $C_{\a}^A \otimes C_{\b}^B $.
The decoherence functional for the composite system factors,
\beq
D^{AB} (\a, \b, \a', \b') = D^A (\a, \a') D^B (\b ,\b')
\eeq
From this we see that the condition of decoherence passes the Di\'osi test, but
consistency fails -- the requirement Eq.(\ref{1.13}) on the subsystems does not
imply the same requirement on the composite system because
\beq
{\rm Re} D^{AB} = ({\rm Re} D^A) ({\rm Re} D^B) - ({\rm Im} D^A)   ({\rm Im} D^B)
\eeq
Similarly, we have that
\beq
q^{AB} (\a, \b) = q^{A} (\a) q^{B} (\b )
\eeq
so the linear positivity condition Eq.(\ref{1.22}) on subsystems does not
imply linear positivity on composite system.
This clearly rules out consistency and linear positivity
as reasonable conditions for the assignment of probabilities to histories \cite{Dio2}.

Now the important question is whether partial decoherence satisfies the Di\'osi condition.
For a composite system, both $p^{AB}(\a, \b)$ and $q^{AB}(\a, \b)$ factor, and our condition
Eq.(\ref{1.20}) reads
\beq
p^A (\a) p^B (\b) = q^A (\a) q^B (\b)
\label{1.29}
\eeq
This condition will clearly hold if Eq.(\ref{1.20}) holds for each subsystem so
the Di\'osi condition is satisfied.

Given that consistency fails the Di\'osi test and decoherence passes it, one might have
thought that a condition satisfying the Di\'osi test must necessarily be {\it stronger}
than consistency, but partial decoherence disproves this idea, being neither stronger
nor weaker than it, just different.

\section{A Strengthened Di\'osi Test}

The Di\'osi Test, as stated above, is sufficient to rule out linear positivity and
consistency, but does not rule out partial decoherence. This state of affairs
may be satisfactory, but it is of interest to revisit the Di\'osi test and ask
whether a strengthened version should be considered. In its original statement, the logical implication in the Di\'osi test goes in one direction only: the condition on subsystems must imply
the same condition on the composite system. But should the condition also satisfy
a similar test with the reverse implication? We refer to such a test as:

\noindent{\bf The Reverse Di\'osi Test:} {\it Suppose a composite system consists of a number of statistically
independent subsystems. Then any condition for the assignment of the probabilities
applied to the composite system must imply the same condition for
each subsystem.}

It seems reasonable to require that any probability assignment condition should
satisfy both Di\'osi tests.  The argument for the reverse test is quite different
to the original one. It is actually about
coarse graining and is essentially the requirement that any conditions for the assignment
of probabilities to histories must be preserved in form under coarse grainings.

It is easy to see that decoherence passes the Reverse Di\'osi Test, but consistency and
linear positivity fail it. What about partial decoherence?
Does Eq.(\ref{1.29}) imply Eq.(\ref{1.20}) for each subsystem? This is more subtle
than the previous cases.
Summing over $\b$
we obtain
\beq
p^A (\a) \sum_{\b} p^B (\b) = q^A (\a)
\label{1.30}
\eeq
We now see that the desired result depends on whether the histories are homogeneous
or inhomogeneous. For homogeneous histories
\beq
\sum_{\b} p^B (\b) = 1
\label{1.31}
\eeq
and it follows that $ p^A (\a ) = q^A (\a) $ (and similarly for $B$), as desired.

For inhomogenous histories, we sum Eq.(\ref{1.30}) over $\a$, to obtain
\beq
\sum_\a p^A (\a) \sum_{\b} p^B (\b) = 1
\label{1.32}
\eeq
If $A$ consists of homogenous histories, then its probabilities sum to $1$
which forces the histories of $B$ to sum to $1$. The Di\'osi condition is then satisfied.
However, if both $A$ and $B$ consist of inhomogeneous histories, then Eq.(\ref{1.32})
may be satisfied without the probabilities of either system summing to $1$ (and recall
that the probabilities do not necessarily need to sum to $1$ for inhomogeneous histories).
Eqs.(\ref{1.32}) and (\ref{1.30}) may be combined to read
\beq
 q^A (\a) = \frac { p^A (\a) } { \sum_{\a'} p^A (\a') }
\eeq
This is a consistent relationship between $q$ and $p$ but narrowly
falls short of satisfying the condition of partial decoherence for
the subsystem, so the Reverse Di\'osi Test is not satisfied in this case. The heart
of the difficulty here is that for inhomogenous histories and statistically
independent subsystems, the relation Eq.(\ref{1.32})
does not imply that the subsystem probabilities sum to $1$.

One could contemplate requiring that the Reverse Di\'osi Test is restricted
to identical or near-identical susbystems. This would mean that the subsystem
probabilities must sum to the same (or almost the same) value, which could
be greater than $1$ or less than $1$, but Eq.(\ref{1.32}) would then
force both probabilities to sum to $1$. But there is no obvious reason
for restricting to near-identical susbystems.

The failure of partial decoherence to fully pass this test hinges around the
unusual properties of inhomogeneous histories. At present no general statements
are known about the sums of their probabilities (in the absence of decoherence).
That is, one would like to know whether $ \sum p \ne 1 $ is a generic feature
or one that can happen only in exceptional circumstances. This part of the story
remains unclear.

\section{Another Test: Robustness under Change of Dynamics}

Di\'osi also considered another test that any probability assignment condition
ought to satisfy \cite{Dio}. Suppose we apply some sort of external field to
the physical system so producing a perturbation in the Hamiltonian. One can imagine
that such a perturbation could be chosen to be essentially classical -- i.e., in such
a way that it does not introduce extra quantum coherence. Clearly any probability assignment
condition should be robust under such a perturbation, meaning that it should not change in form.

Di\'osi gave a specific example of such
a perturbation acting at just one moment of time $t_k$
and argued that it produces a change in the class operators of the form
\beq
C_{\a} \rightarrow e^{ - i \lambda_{\a_k} } U_k^{\dag} C_{\a}
\label{C}
\eeq
where $\lambda_{\a_k}$ is a real number and $U_k$ is a unitary operator whose exact form
is not required here. The decoherence functional then changes according to
\beq
D(\a, \a') \rightarrow e^{ i ( \lambda_{\a_k} - \lambda_{\a_{k}'} )} \ D( \a, \a')
\eeq
Clearly the condition of decoherence is preserved under this transformation, but consistency
and linear positivity are not. Partial decoherence also appears to fail this test since $p(\a)$
is preserved but $q(\a)$ changes.

This test does not appear to have been investigated much beyond the few simple observations
made here and its status is less clear than the other tests described above. It seems
to be related to the general idea that the decoherent histories approach concerns
the question of determining those situations to which to probabilities can be assigned
independently of whether the system is actually measured. Indeed the specific example
of a perturbation given above is exactly of the form of a physical measurement.

One wonders whether both this test and the previous tests involving
subsystems are examples of a more general set of requirements concerning classicality-preserving
operations which one would expect any probability assignment condition to satisfy.
This would be of interest to investigate in the future.

\section{Discussion}

The account of decoherent histories presented here confirms decoherence, diagonality of the decoherence
functional, as the only sensible condition for the assignment of probabilities
for histories. Partial decoherence comes close, but fails in some subtle
ways related to the properties of inhomogeneous histories. Consistency and
linear positivity fail very clearly to be satisfactory.

Decoherence has a nice geometric picture in that (for pure states) it
corresponds to orthogonality of the set of states $ \{ C_\a | \psi \rangle \} $.
The robustness test particularly recommends this picture since the transformation
Eq.(\ref{C}) implies a unitary transformation on the states $C_{\a} | \psi \rangle$
under which all orthogonality properties are preserved.
To see why the orthogonality of these states should be important one needs
to look at the underlying mechanisms which cause any of the probability assignment
conditions to become satisfied. There are essentially two such mechanisms.

The first relates to conservation, either in terms of a system-environment
split (where one system is much slower than the other, so its variables
are approximately conserved), or in terms of local densities, which are
approximately conserved when averaged over large volumes. When there is decoherence
due to conservation, the decoherence of histories comes about essentially because
there exist certain states which are preserved in form by the action of the class
operator $C_{\a}$, and thus the set of states of the form $C_{\a} | \psi \rangle $ are orthogonal \cite{Hal5}.

The second mechanism is to do with statistics (and also arises in the situation
where there is a system environment split, but at much finer-grained scales
than the conservation situation described above). The point here is that for large systems,
any pair of ``typical'' states will tend to be approximately orthogonal.

In both cases, therefore, the mechanisms producing decoherence refer to
orthogonality of states, which is why decoherence functional diagonality has such a central role.

There are however, certain issues that remain incompletely understood. As stated earlier,
the physical mechanisms that cause the probability assignment conditions to become
satisfied tend to produce decoherence, and not just partial decoherence or consistency.
This actually means that, if there is a physical mechanism present such
as conservation or an environment, then it might be sufficient {\it in practice} to check
only that one of the weaker conditions holds, such as partial decoherence, since the above
argument suggests that full decoherence will then probably hold too. Indeed, one of the motives for
investigating partial decoherence is that it is in practice much simpler to check than
the diagonality of the whole decoherence functional. It would be of interest to make these vague
ideas more precise. For example, one wonders if it is possible to write down a simple
auxiliary condition which signifies in a general way the presence of a
physical decoherence mechanism, such as conservation or an environment, but without explicitly
identifying the mechanism. Such an auxiliary condition, adjoined to partial decoherence
or consistency might then be equivalent to full decoherence. Differently put, the question
is the following: can the condition
of decoherence be split into two conditions in a useful way: partial decoherence (or consistency), plus
some other auxiliary condition reflecting the underlying physical mechanism?
(The existence of records is an example
of a possible auxiliary condition but this is probably too strong for what is being suggested here).
These ideas will be pursued elsewhere.

\section{Acknowledgements}

I am very grateful to Lajos Di\'osi, Fay Dowker, Jim Hartle and James Yearsley for useful conversations
and comments on the manuscript.
I would also like to thank Ting Yu and Bei-Lok Hu for inviting me to contribute
to this volume.

\appendix

\section{A Situation Exhibiting Consistency but not Decoherence}

The following is a simple example of a situation in which there is exact consistency
but the imaginary part of the decoherence functional is non-zero so there is no decoherence.
We consider at system which has two alternatives at each moment of time denoted by the projectors
$P$ and $\bar P = 1 - P $. A useful example to visualize is the case of a point particle
with projections onto the positive or negative $x$-axis.

We consider histories characterized by alternatives at two moments of time,
$t_1$, $t_2$, and we introduce Heisenberg picture projectors,
\beq
P_1 = P (t_1), \ \ \ \ P_2 = P(t_2)
\eeq
Then we consider a pair of (inhomogeneous) histories represented by the class operators
\beq
C = P_2 P_1 + \bar P_2 \bar P_1
\eeq
and
\beq
\bar C = 1 - C = P_2 \bar P_1 + \bar P_2 P_1
\eeq
In the case of projections onto the positive and negative $x$-axis,
$ C $ represents the statement that the particle is on the same side
of $ x = 0 $ at both $t_1 $ and $t_2 $. $\bar C $ represents the
statement that the particle at time $t_2$ is on
the side of $x=0$ opposite to that it was on at $t_1$.

The real part of the off-diagonal term of the decoherence functional
is given by,
\bea
2 {\rm Re} D &=& {\rm Tr} \left( C \rho \bar C^{\dag} \right)
+ {\rm Tr} \left( \bar C \rho C^{\dag} \right)
\nonumber \\
&=& {\rm Tr} \left( ( \bar C^{\dag} C + C^{\dag} \bar C) \rho \right)
\eea
It is easy to see that
\bea
\bar C^{\dag} C + C^{\dag} \bar C  &=& P_1 P_2 \bar P_1 +
\bar P_1 \bar P_2 P_1 + \bar P_1 P_2 P_1 + P_1 \bar P_2 \bar P_1
\nonumber \\
&=& P_1 \bar P_1 + \bar P_1 P_1 = 0
\eea
Hence, ${\rm Re D} = 0 $, although ${\rm Im} D $ is generally
non-zero. The set of histories is therefore exactly
consistent for any initial state but generally not decoherent.

The significance of this example is not clear, although the class operator
$\bar C$ gives a crude semiclassical description of crossing the origin
during a given time interval, and as such may be relevant to the decoherent
histories analysis of the arrival time problem \cite{HaYe}. This will be investigated elsewhere.

\bibliography{apssamp}

\end{document}